\documentclass[conference]{IEEEtran}
\usepackage{cite}

\ifCLASSINFOpdf
  \usepackage[pdftex]{graphicx}
\else
\fi
\usepackage{algorithmic}

\usepackage{array}

\ifCLASSOPTIONcompsoc
\usepackage[caption=false,font=normalsize,labelfont=sf,textfont=sf]{subfig}
\else
  \usepackage[caption=false,font=footnotesize]{subfig}
\fi

\ifCLASSOPTIONcaptionsoff
 \usepackage[nomarkers]{endfloat}
 \usepackage{authblk}
\let\MYoriglatexcaption\caption
\renewcommand{\caption}[2][\relax]{\MYoriglatexcaption[#2]{#2}}
\fi
\usepackage{url}

\usepackage{amsmath}
\usepackage{verbatim}

\usepackage{footnote}
\makesavenoteenv{tabular}

\usepackage{multirow}
\usepackage{graphicx}
\usepackage{listings}
\usepackage{xcolor}

\usepackage{fancyvrb}
\usepackage{subcaption}
\usepackage[ruled,vlined]{algorithm2e}

\lstdefinelanguage{json}{
    basicstyle=\normalfont\ttfamily,
    numbers=left,
    numberstyle=\scriptsize,
    stepnumber=1,
    numbersep=8pt,
    showstringspaces=false,
    breaklines=true,
    frame=lines,
    backgroundcolor=\color{background},
    literate=
     *{0}{{{\color{numb}0}}}{1}
      {1}{{{\color{numb}1}}}{1}
      {2}{{{\color{numb}2}}}{1}
      {3}{{{\color{numb}3}}}{1}
      {4}{{{\color{numb}4}}}{1}
      {5}{{{\color{numb}5}}}{1}
      {6}{{{\color{numb}6}}}{1}
      {7}{{{\color{numb}7}}}{1}
      {8}{{{\color{numb}8}}}{1}
      {9}{{{\color{numb}9}}}{1}
      {:}{{{\color{punct}{:}}}}{1}
      {,}{{{\color{punct}{,}}}}{1}
      {\{}{{{\color{delim}{\{}}}}{1}
      {\}}{{{\color{delim}{\}}}}}{1}
      {[}{{{\color{delim}{[}}}}{1}
      {]}{{{\color{delim}{]}}}}{1},
}

\begin{document}
%
\title{A Scalable In Transit Solution for \\ Comprehensive Exploration of Simulation Data}

\title{SeerX: A Dynamic Resource Allocation In situ  Framework with Data Reduction}

\author{
\IEEEauthorblockN{Pascal Grosset}
\IEEEauthorblockA{Los Alamos National Laboratory\\
Email: pascalgrosset@lanl.gov}
\and
\IEEEauthorblockN{James Ahrens}
\IEEEauthorblockA{Los Alamos National Laboratory\\
Email: ahrens@lanl.gov}
}

\markboth{Journal of \LaTeX\ Class Files,~Vol.~14, No.~8, August~2015}%
{Shell \MakeLowercase{\textit{et al.}}: Bare Demo of IEEEtran.cls for IEEE Journals}
%



\maketitle

\begin{abstract}


As simulations produce more data than available disk space on supercomputers, many simulations are employing in situ analysis and visualization to reduce the amount of data that needs to be stored. While in situ visualization offers potential for substantial data reduction, its efficacy is hindered by the need for a priori knowledge. First, we need to know what  visualization parameters to use to highlight features of interest. Second, we do not know ahead of time how much resources will be needed to run the in situ workflows, e.g. how many compute nodes  will be needed for in situ work. In this work, we present SeerX, a lightweight, scalable in-transit in situ service that supports dynamic resource allocation and lossy compression of 3D simulation data. SeerX enables multiple simulations to offload analysis to a shared, elastic service infrastructure without MPI synchronization.

\end{abstract}


%
\IEEEpeerreviewmaketitle

\section{Introduction}

As the computational power of supercomputers continues to grow, large-scale simulations are producing data at rates that exceed the capacity of high-performance computing (HPC) storage systems. To mitigate this challenge, in situ analysis and visualization techniques are increasingly employed to extract insights while the simulation is running, thereby significantly reducing the volume of data that must be written to disk. In response to this need, several frameworks have been developed to support in situ workflows, including Ascent~\cite{Ascent:2017}, Catalyst~\cite{Ayachit:2021}, and Sensei~\cite{Bethel:2022}.

\begin{figure}[b]
\vspace{-0.3in}
\centering 
\subfloat[Front View]{
    \includegraphics[width=0.49\linewidth]{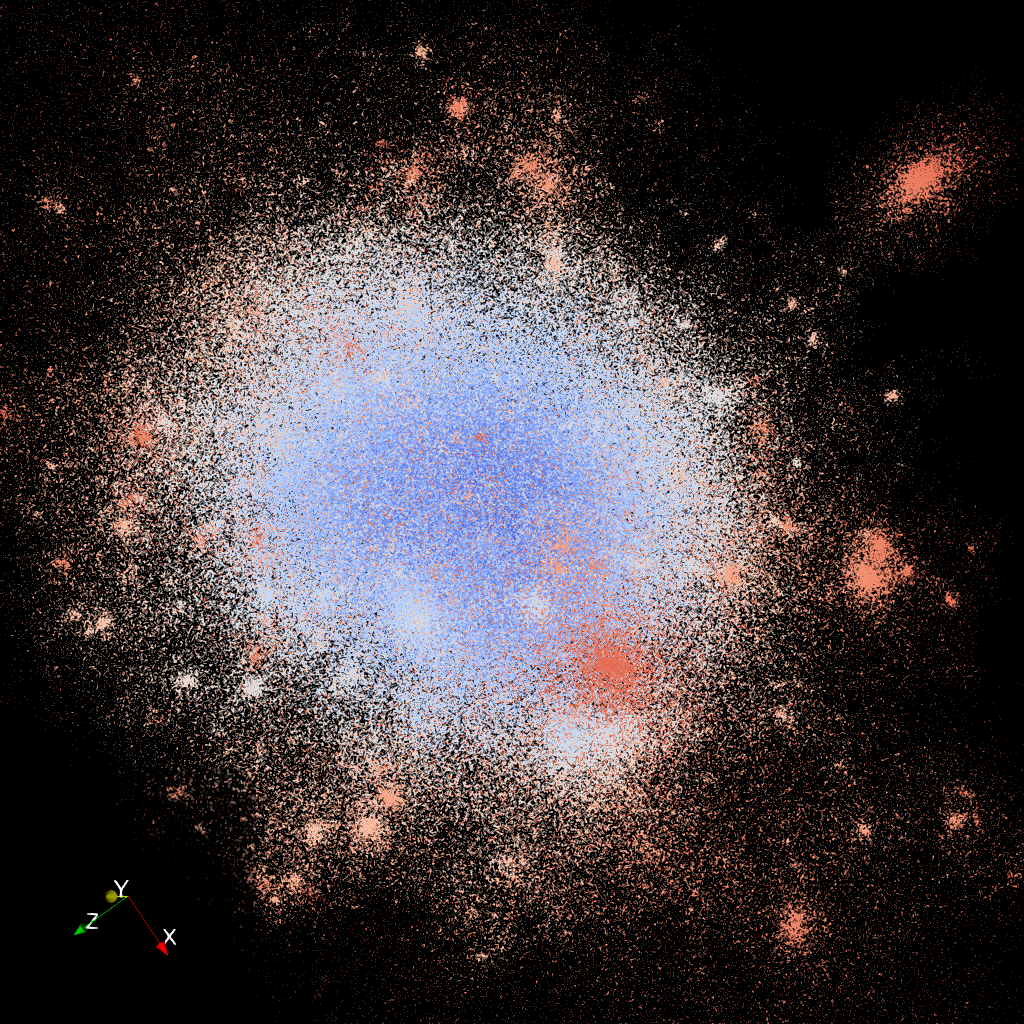} 
} 
\subfloat[Side view]{
    \includegraphics[width=0.49\linewidth]{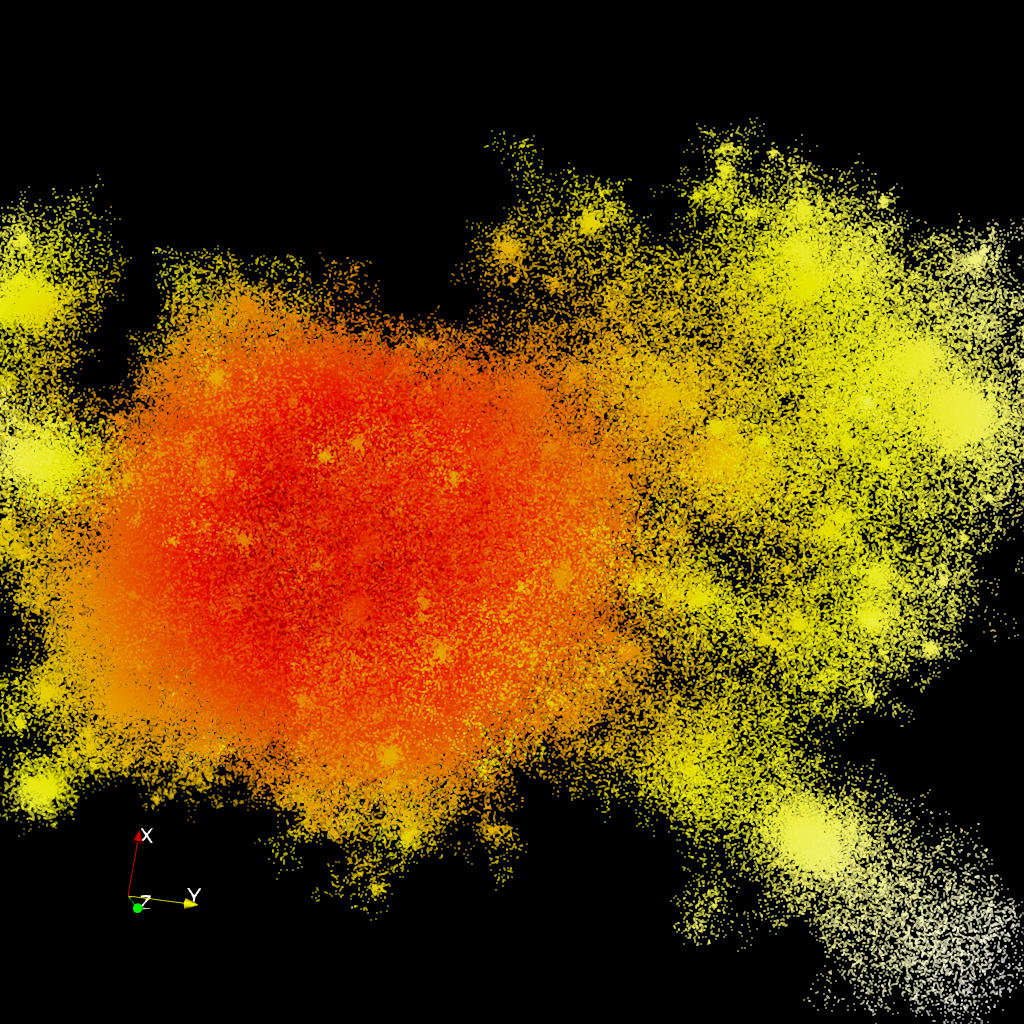} 
} \hfil 
\centering
\caption{Halo (cluster of particles) in the HACC cosmology simulation viewed from different angles using different color maps. In (b) we see that the large halo has many small halos attached to it.} 
\label{fig:insitu} 
\end{figure}

One challenge with in situ visualization is that users must decide in advance which features to visualize and how to visualize them, including choices such as transfer functions, camera angles, and colormaps. For example, Figure~\ref{fig:insitu} shows different renderings of the same halo form a HACC cosmology dataset. As can be seen, varying the colormap and camera position can result in significantly different views of the same feature. In other words, depending on the camera position, we might be missing out on some very important features.

Visualization techniques can be grouped into three categories: analytical, exploratory, and production~\cite{Butler:1993}. Analytical visualization focuses on predefined representations; exploratory visualization supports interactive data exploration to reveal patterns or anomalies; and production visualization aims to generate high-quality imagery for dissemination. While current in situ tools are effective for analytical and production visualization, they are less suited to exploratory analysis due to their static and non-interactive nature. One strategy to address this limitation is to pre-compute a range of visualizations and store them in a Cinema database~\cite{ahrens2014image}, but this can result in a prohibitively large number of images~\cite{Bauer:2016}, consuming significant storage space.

An alternative approach to reducing storage demands is data compression. Several error-bounded lossy compression algorithms, such as SPERR~\cite{Li:2023}, TTHRESH~\cite{ BLP:19}, and SZ3~\cite{sz3} can dramatically reduce data size while guaranteeing error bounds specified as absolute error, relative error, and PSNR. Moreover, recent work by Etchi et al.~\cite{etchi:2024} and Wang et al.~\cite{Wang:2022} have demonstrated the scientific visualization on compressed data can maintain very high quality even after substantial size reduction. Leveraging these compressors can thus allow us to achieve data reduction while drastically reducing the amount of data that needs to be preserved, in line with the goal of in situ visualization.

In situ tools have traditionally shared the same resources (in-line) as simulations. This tight coupling introduces resource contention and can degrade performance, especially at scale~\cite{kress:2020}. In contrast, in-transit approaches decouple analysis from simulation by offloading computation to separate resources. Systems like SERVIZ~\cite{Ramesh:2022} have demonstrated performance improvements of up to 26\% over in-line methods. However, current in-transit architectures typically rely on static resource allocation.

To address these limitations, we introduce \textbf{SeerX} (Seer e\textbf{X}tended), a lightweight, scalable in-transit \textit{in situ} framework designed to dynamically adapt to changing analysis workloads. SeerX offers the following key features:
\begin{itemize}
    \item Elastic resource scaling: dynamic allocation and deallocation of resources for analysis
    \item Integrated data reduction: support for lossy compression of simulation data during runtime
    \item MPI decoupling: no requirement for MPI synchronization, enabling asynchronous operation
\end{itemize}

To the best of our knowledge, SeerX is the first system to provide a fully scalable in-transit in situ infrastructure that treats data reduction as a first-class capability.

In addition, SeerX provides compatibility with existing analysis libraries, such as ParaView Catalyst, through a simple API that can be invoked from MPI programs. By shifting synchronization responsibilities to the analysis service, SeerX removes the need for MPI communicators in the simulation, making it suitable for integration with asynchronous task-based runtimes, such as Legion~\cite{10.5555/2388996.2389086}.

\section{Related Work}

\subsection{In Situ Frameworks}
A number of \textit{in situ} frameworks have been developed to support large-scale simulation workflows. Two of the most widely used are ParaView Catalyst~\cite{Catalyst:2015} and Ascent~\cite{Ascent:2017}. Ascent is a lightweight \textit{in situ} library offering flexible visualization capabilities, including Jupyter-based analysis. Catalyst 2~\cite{Ayachit:2021}, the successor to ParaView Catalyst, introduces a generalized interface supporting multiple backends, including both ParaView and Ascent. However, both Catalyst and Ascent operate in-line, meaning they share resources with the simulation and do not natively support data reduction techniques.

To address these limitations, Mazen et al.~\cite{mazen2024} combined Catalyst 2 with ADIOS2~\cite{godoy_adios} to support both in-line and in-transit analysis. Their approach transfers data artifacts to analysis nodes, incorporating limited data reduction. However, they do not transmit full 3D datasets, nor do they evaluate compression algorithms for scalable visualization.

SERVIZ~\cite{Ramesh:2022} is a shared in situ service architecture that supports visualization across multiple simulations. While SERVIZ offloads analysis to dedicated nodes, it uses a fixed allocation size and cannot dynamically adjust the number of resources. More recently, Dorier et al. proposed Colza~\cite{Dorier:2022, DORIER2023106}, which supports elastic resource management by allowing the number of MPI ranks in a job to be modified at runtime. To overcome MPI's rigidity, they rely on dependency injection (e.g., replacing MPI with MoNA~\cite{mona} in VTK) or dependency overload (e.g., intercepting MPI calls in VTK-h). Although this enables elasticity, the integration process can be complex, but they do achieve substantial performance improvements, up to 40 seconds in rendering time over static configurations, which demonstrate the value of elastic resource scaling.

Both SERVIZ and Colza use components from the Mochi project to implement their services. In contrast, our proposed system also leverages Mochi (see Section~\ref{mochi}), but differs by using independently allocated nodes that communicate via TCP, avoiding the need for MPI and enabling seamless cross-allocation communication. This allows SeerX to \textit{dynamically scale analysis resources} without strict synchronization or static reservations.

Our framework builds upon previous work in Seer~\cite{Grosset_isav:2020} and Seer-Dash~\cite{Grosset_isav:2021}. Seer enables personalized \textit{in situ} steering, while Seer-Dash introduces point sampling and a dashboard-based interface for visual exploration. Both rely on Mochi key-value stores but use them only for steering metadata, not for storing simulation data. Additionally, Seer-Dash’s data reduction is limited to particle sampling, which is applicable only to Lagrangian simulations. Many simulations, however, are Eulerian, where spatial sampling is not effective—highlighting the need for general-purpose data reduction approaches.

\subsection{Data Reduction}

Lossless compression algorithms such as LZ4, ZLib, and Zstandard are widely used in HPC settings, but they typically achieve only 2X–3X compression on scientific floating-point data~\cite{Zeyen:2017}. To address this, several error-bounded lossy compressors have been developed, including ZFP~\cite{lindstrom2014fixed}, SZ3~\cite{sz3}, MGARD~\cite{Ainsworth2018}, and TTHRESH~\cite{BLP:19}. These algorithms provide significantly high compression ratios and allow users to tune compression fidelity depending on the sensitivity of downstream analysis or visualization tasks.
Moreover, recent work in deep learning are continuously improving the compression ratio while decreasing the training time. Di et al. have an excellent survey of compression algorithms applied to scientific datasets~\cite{di_survey:2024}.

For this project, we incorporated BLOSC~\cite{blosc2024} for lossless compression and SZ3 for lossy compression. BLOSC is a meta-compressor that supports multiple backends including LZ4, Zlib, and Zstandard. We use FastLZ~\cite{fastlz} as the default, but other algorithms can be easily configured. SZ3 offers multiple operational modes (e.g., Lorenzo, regression), each providing different trade-offs between compression speed and accuracy. It supports all standard error bounding schemes and offers both C and Python APIs, making it suitable for integration into heterogeneous workflows. Our choice of BLOSC and SZ3 is guided by their efficiency, configurability, and portability across both compiled and scripting environments (they both provide a C and Python API). This allows SeerX to perform on-the-fly compression of 3D simulation data before offloading it to a scalable in-transit analysis infrastructure.
\section{Architecture}

\begin{figure*}[htb!]
    \centering
    \includegraphics[width=0.99\linewidth]{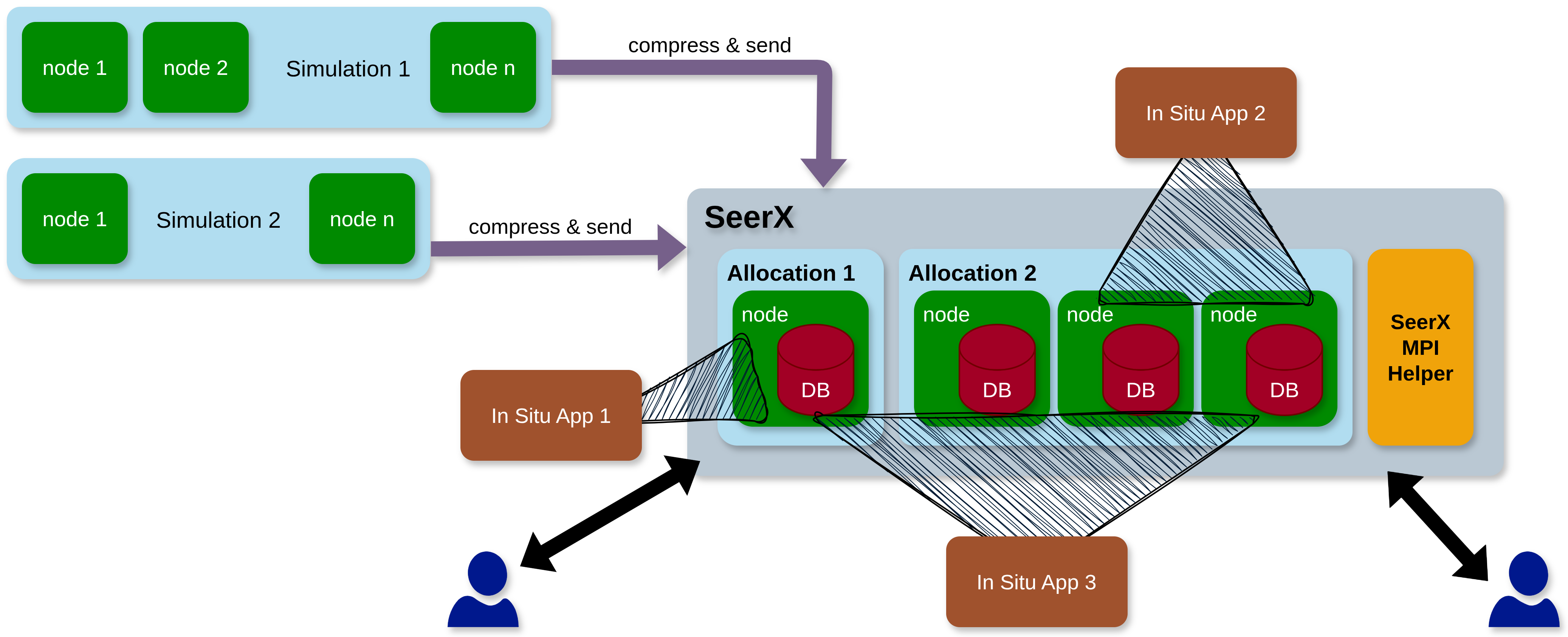}
    \caption{Overview of the SeerX architecture with  two simulations are connected and sending data to the service. The SeerX service can span multiple allocations (two in this case) and in situ applications can connect to multiple nodes with databases as needed. Users can also connect to SeerX in real time as well, through the Trame to interact with the data.}
    \label{fig:architecture}
    \vspace{-0.15in}
\end{figure*}

Figure~\ref{fig:architecture} provides a high-level overview of SeerX. At specified timesteps, simulation nodes compress and send their data to key-value databases managed by a Mochi server. Users can subsequently access this data directly, download it for offline analysis, and interact with it through existing in situ visualization libraries using the SeerX MPI helper.

\subsection{Elastic Resource Scaling}\label{mochi}
The SeerX architecture leverages the Mochi framework, a modular toolkit for building distributed HPC data services. Mochi consists of several core modules: \textit{Mercury}, an HPC-oriented Remote Procedure Call (RPC) library; \textit{Argobots}, a flexible threading and tasking runtime; \textit{Margo}, a distributed service layer combining Mercury and Argobots; and \textit{Yokan}, a key-value storage service supporting multiple backends, including LevelDB and BerkeleyDB. Mochi also provides \textit{Bedrock}, a bootstrapping framework for initializing and managing services.

In SeerX, we specifically use the Yokan in-memory key-value store. Each Yokan database instance is initialized using Bedrock, which assigns it an IP address and port for RPC access. Unlike traditional HPC environments, where communication is typically restricted within a single node allocation, the \textbf{key idea} here is \textbf{Mochi’s RPC-based approach via TCP allows nodes from different allocations to communicate seamlessly}. This, in turn allows us to add/remove resources on the fly thereby enabling elasticity scaling  on workload fluctuations.

\subsection{Integrated Data Reduction}
SeerX integrates both lossless and lossy compression methods through BLOSC and SZ3. Users specify the desired compression methods and parameters using a straightforward JSON configuration (Listing~\ref{lst:compression}). For example, particle IDs might require lossless compression via BLOSC, whereas particle positions might tolerate lossy compression via SZ3. Moreover, compression settings can be dynamically adjusted for different simulation timesteps or variables, enhancing flexibility and data management efficiency.

\begin{lstlisting}[language=sh, caption=Compression specification for different fields of a simulation,basicstyle=\small\ttfamily,label=lst:compression]
"data" : [
    {
        "name": "id",
        "compressor": "BLOSC" },
    {
        "name": "x",
        "compressor": "SZ3",
        "mode": "abs",
        "value": 0.003 },
          . . .  ]
\end{lstlisting}
\vspace{-0.15in}

\subsection{Data Access}
Data stored in the Yokan databases can be accessed through the Mochi Yokan Python API or via our lightweight Python helper library. Users can interactively explore the data within a Jupyter notebook and visualize it using frameworks such as Trame~\cite{Avery:2024}, a VTK-based Python visualization toolkit. Additionally, data can be downloaded, decompressed, and further analyzed using standard visualization tools like ParaView.


To support integration with existing workflows based on libraries such as Ascent and Catalyst, SeerX provides the \textit{SeerX MPI Helper} (shown as the blue box in Figure~\ref{fig:architecture}). The SeerX MPI Helper is a C++ API offering straightforward functions for retrieving and decompressing stored data and metadata from the Mochi databases. For example, a simple function call like: $getData(int~timestep, int~rank, std::string~variable, size\_t~\&n, std::string~\&metadata)$ allows easy extraction and subsequent passing of the data to standard in situ libraries. The SeerX MPI Helper also keeps track of which timesteps are ready and supplies this to the in situ backend as needed. Although using this helper requires allocating analysis nodes, it significantly reduces overhead compared to running visualization in-line with the simulation. At scale, in-line visualization often encounters bottlenecks due to image compositing and synchronization overhead~\cite{Grosset:2016}. Figure~\ref{fig:architecture} also shows how 3 applications could connect to the in situ service and span different allocations.

\subsection{The SeerX Workflow}

The integration of SeerX into simulation code is straightforward. The first step is to allocate one or more nodes and launch Mochi Yokan databases. The address and port of where these servers are spun up are collected. For each simulation, a JSON config file is constructed as shown in Code listing 2.

\begin{lstlisting}[caption=Simulation Configuration File,label=lst:sim,basicstyle=\small\ttfamily]
{ "sim-id" : "XXXX",
  "libraries" : "<path to libyokan-bedrock>
  "providers": [
     {
       "name" : "yokan_provider",
       "provider_id": 124,
       <mochi-configuration params ...>
     } ],
  "databases" : [
    {
      "address": "192.168.81.72:46593",
      "protocol": "ofi+tcp"
    },
    {
      "address": "192.168.81.78:45217",
      "protocol": "ofi+tcp"
    } ],
  "data": <see listing 1>      }
\end{lstlisting}

Each simulation configuration file has a unique simulation identified (sim-id) and lists the database it intends to use. Different simulations can use different databases as needed; e.g. a large simulation might need 4 nodes while a small one needs only one. Integration in the simulation code is as shown in Code listing~\ref{lst:integration}. As mentioned before, we do not require an MPI communicator, thereby enabling complete \textbf{MPI decoupling}. So, integration into the simulation requires only specifying the rank id (or task ID in an asynchronous tasks-based environment) and sending the data. If multiple nodes are being used for the in situ services, the data from the simulation (e.g. temperature, pressure) is then evenly distributed across the nodes, while the metadata (e.g. data type of the field temperature, the number of elements for that rank) is stored on Yokan node 0 to facilitate retrieval. At any time, additional Yokan nodes can be spun up and down for the service as needed. The only requirement is to spin up the databases before a simulation starts. 

\begin{lstlisting}[language=C++, caption=Simulation Code Integration,basicstyle=\small\ttfamily,label=lst:integration]
InSitu seerx; // initialize seer
seerx.init(num_ranks, "mochi-yokan-config.json");
    . . .
    
for (ts=0; ts< numTs; ts++){ // sim loop
  // compute for simulation
         .  .  .
             
  // send data
    seerx.sendData(myrank, step, "x", 
       "data", "float", nTotal, &x[0]);
    seerx.sendData(myrank, step, "y", 
       "data", "float", nTotal, &y[0]);
                .  .  .
    seerx.tsDone(myrank, ts);   }
seerx.simDone(myrank, ts);
\end{lstlisting}
\section{Results and Discussion}
We evaluated SeerX on the Darwin HPC cluster at Los Alamos National Laboratory, utilizing its scaling partition with 64 compute nodes, each equipped with dual Intel Broadwell E5-2695\_v4 CPUs (36 cores per node, 125 GB memory) and interconnected via a 100Gb Infiniband EDR network. The performance and scalability experiments were conducted using the HACC cosmology simulation code~\cite{Habib:2016}, an extreme scale Lagrangian cosmology solver employing a hybrid gravity solver and Conservative Reproducing Kernel Smoothed Particle Hydrodynamics (CRK-SPH) for gas dynamics.

\begin{figure}[h]
\vspace{-0.2in}
    \centering
    \includegraphics[width=1.0\linewidth]{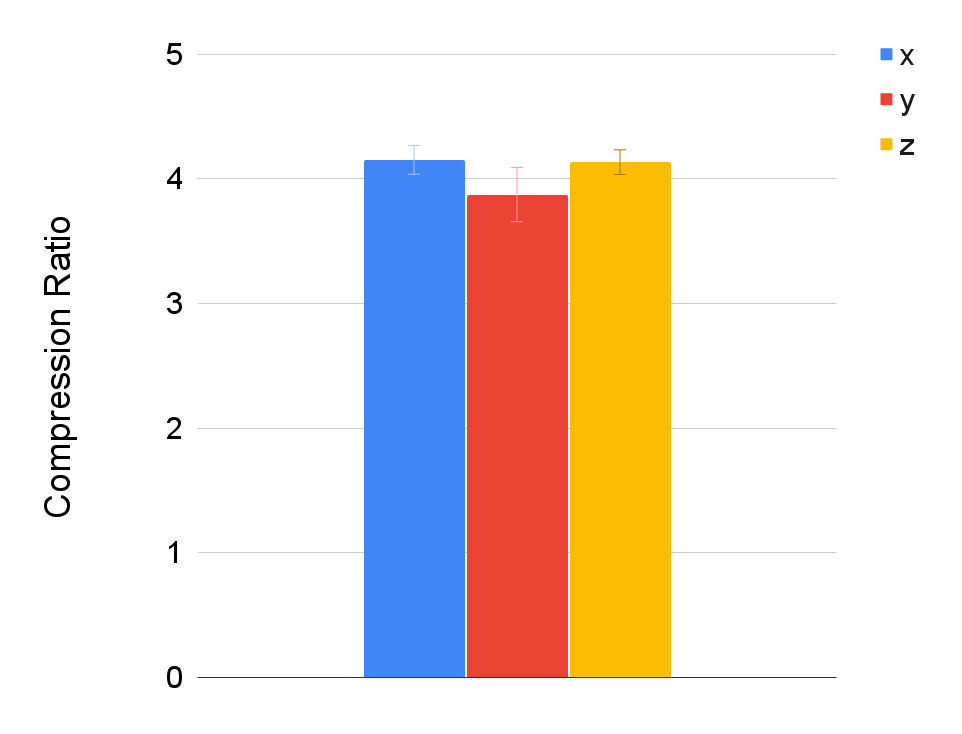}
    \vspace{-0.3in}
    \caption{Compression ratios achieved with SZ3 for particle coordinates ($x$, $y$, $z$) using an absolute error bound of 0.003. }
    \label{fig:compression-ratio}
\end{figure}

\begin{figure}[b!]
    \vspace{-0.25in}
    \centering
    \includegraphics[width=1.0\linewidth]{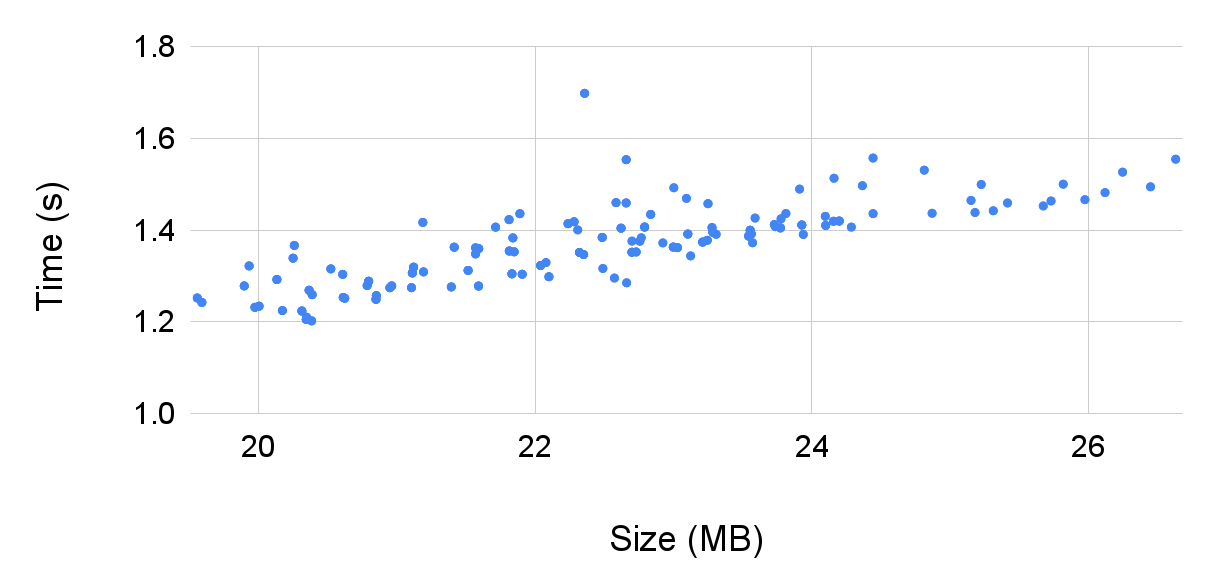}    
    \caption{Measured throughput for compressing HACC data was about 16 MB/s.}
    \label{fig:mochi-throughput}
\end{figure}

\begin{figure}[h]
    \centering
    \vspace{-0.2in}
    \includegraphics[width=1\linewidth]{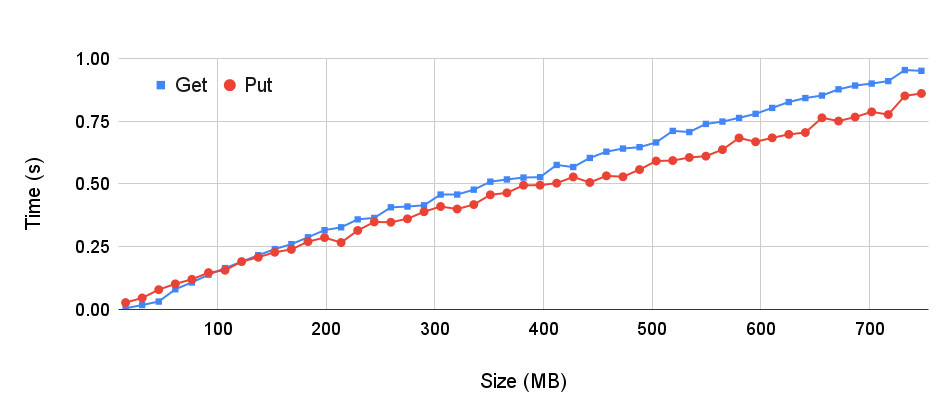}
    \vspace{-0.15in}
    \caption{Scaling test results for data insertion ``Put'', and retrieval ``Get'' performance in Yokan databases. Both operations scale effectively as data sizes increase.}
    \label{fig:data_scaling}
    \vspace{-0.2in}
\end{figure}

For this test, two SeerX database instances were initialized on separate allocations (as shown in Code Listing~2), while the HACC simulation was launched from a third allocation connected to both. Although a full analysis of compression impact on scientific fidelity is beyond the scope of this paper, we configured SeerX to use SZ3 with a maximum absolute error bound of 0.003 for the particle position fields ($x$, $y$, and $z$). This threshold was selected based on prior work by Grosset et al.~\cite{Grosset:2020}, who mentioned that error values of 0.003 are smaller than observational uncertainties in current astronomical instruments.

Figure~\ref{fig:compression-ratio} presents the achieved compression ratios for the $x$, $y$, and $z$ coordinates in a HACC simulation containing 188,758,977 particles. Using the specified SZ3 parameters, we achieved data reduction of about 4X. The average compression throughput was approximately 16~MB/s. By compressing and streaming data directly to SeerX, we significantly reduce I/O overhead and overall storage requirements compared to traditional disk-based workflows. Figure~\ref{fig:mochi-throughput} shows the throughput for compressing the data from HACC.

\begin{figure}[h]
\centering 
    \includegraphics[width=1\linewidth]{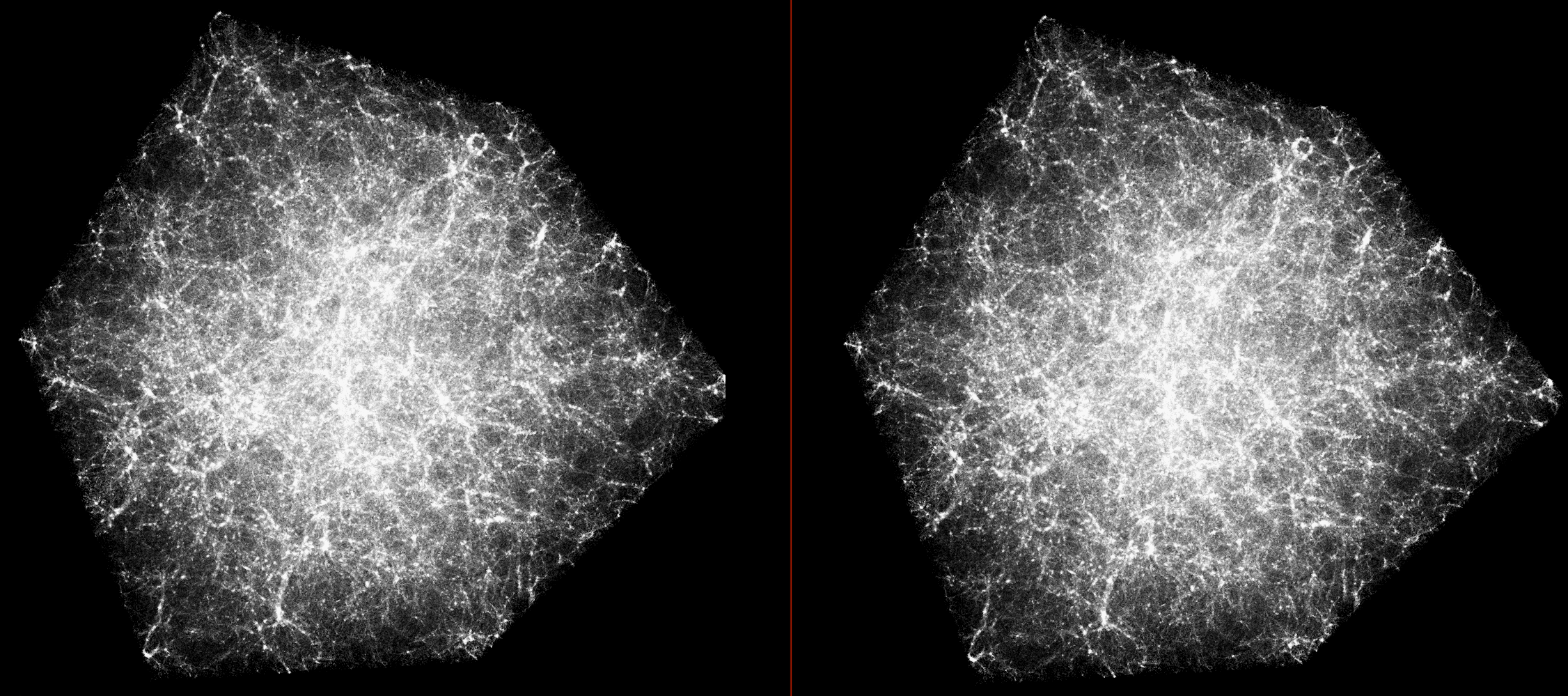}
    \caption{Global comparison of original (left) versus SZ3-compressed (right) HACC simulation data. Compression introduces minimal perceptible differences while significantly reducing dataset size. SSIM score between the images is 0.928.}
    \vspace{0.2in}
    \label{fig:HACC-global}
    \includegraphics[width=1\linewidth]{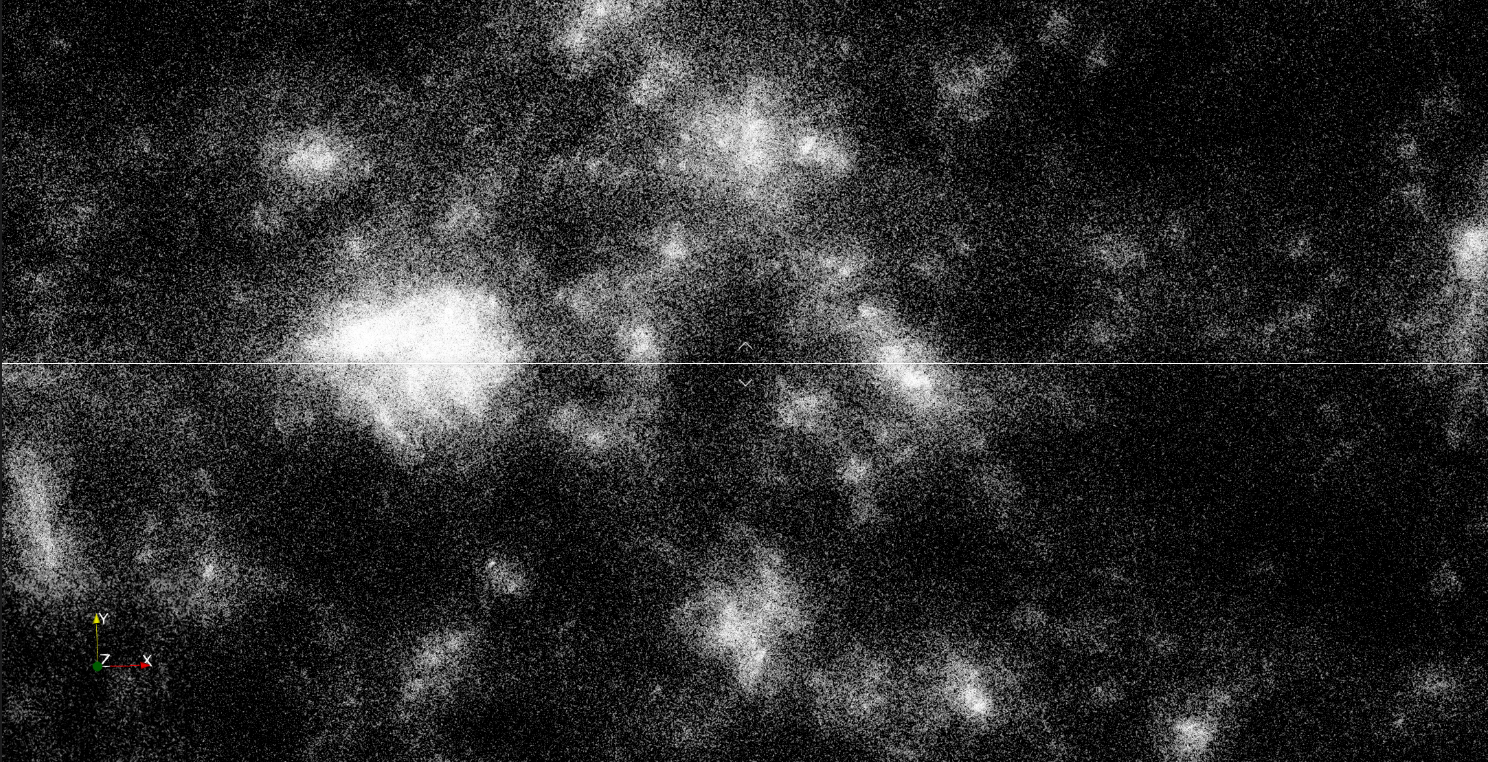}
    \caption{Detailed zoom-in comparison between original (top) and compressed (bottom) data. Despite minor local differences, critical features and structures remain visually identical. SSIM score between the images is 0.895.}
    \label{fig:HACC-local}
    \vspace{-0.2in}
\end{figure}


\begin{figure}[htb!]
    \centering
    \includegraphics[width=0.99\linewidth]{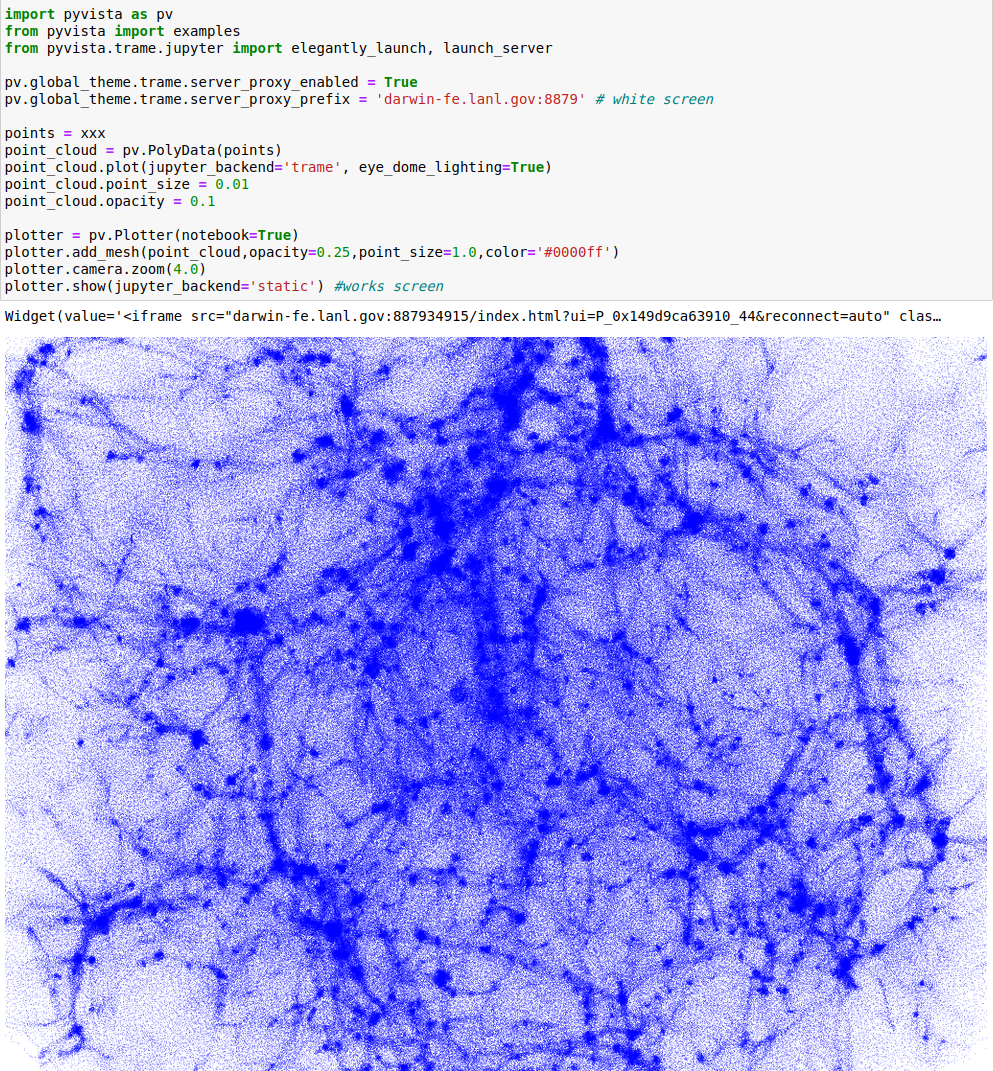}
    \caption{Interactive exploration of compressed data via Jupyter Notebook and Trame visualization. Users can easily zoom, analyze, and interactively query large-scale simulation results.}
    \label{fig:jupyter-trame}
    \vspace{-0.2in}
\end{figure}

To assess scalability, we tested data entry ``Put'' and retrieval ``Get'' performance in the Yokan database as the data size varied. Figure~\ref{fig:data_scaling} shows results of these scalability tests. Both operations exhibit predictable and near-linear scaling behavior, confirming SeerX's ability to efficiently handle increasingly large datasets common in large-scale simulations.

Figures~\ref{fig:HACC-global} and~\ref{fig:HACC-local} present qualitative comparisons between the original (uncompressed) and compressed simulation data using SZ3 (with parameters shown in Figure~\ref{fig:compression-ratio}), along with the corresponding Structural Similarity Index (SSIM) scores; SSIM ranges from 0 to 1, where 1 indicates perfect similarity. Although minor local differences are perceptible, the overall structural fidelity of the dataset is preserved, ensuring that the resulting visualizations remain scientifically interpretable. This is further supported by the SSIM values, which are consistently close to 1, indicating high visual similarity despite the compression.

Once data is stored within Mochi databases, users can retrieve, decompress, and interactively explore the results. Figure~\ref{fig:jupyter-trame} demonstrates this interactive exploration capability, using a Jupyter Notebook server running directly on the analysis nodes and visualizing the data through Trame. The ease and immediacy of such exploration highlight the practical benefit of SeerX's integrated approach to compression and elastic resource allocation.

\section{Conclusion}

In this paper, we presented SeerX, a lightweight and scalable in-transit in situ service. By leveraging the Mochi framework’s TCP-based RPC communication, SeerX supports elastic resource scaling, enabling dynamic addition or removal of compute nodes across multiple independent HPC allocations. We integrated SZ3 and BLOSC compressors for on-the-fly data reduction, enabling efficient storage and comprehensive post hoc exploration of simulation data. Furthermore, by shifting all timestep synchronization to dedicated SeerX service nodes, our approach achieves complete MPI decoupling, providing fully asynchronous in situ analysis compared to traditional frameworks such as ParaView Catalyst and Ascent.
As future work, we aim to evaluate SeerX at larger scales to demonstrate its viability for deployment on full-scale HPC. Additionally, we plan to incorporate automated data redundancy mechanisms, such as periodic backup to persistent storage, and enhance the system’s robustness through dynamic node management to accommodate changing resource demands. 

\section{Acknowledgment}
This work has been authored by employees of Triad National Security, LLC which operates Los Alamos National Laboratory under Contract No. 89233218CNA000001 with  the  U.S.  Department  of  Energy/National  Nuclear  Security  Administration. It was supported by the U.S. Department of Energy, Office of Science, Office of Advanced Scientific Computing Research and Office of DE-SC0021399, Scientific Discovery through Advanced Computing (SciDAC) program. 



\bibliographystyle{IEEEtran}
\bibliography{IEEEabrv,thebib}

\end{document}